\documentclass[aps,pra,reprint,nofootinbib,showkeys]{revtex4-1}
\usepackage{graphicx}
\usepackage{amsmath,amssymb}
\newcommand{\spliter}
{\vskip1ex
 \centerline{\rule{0.6\columnwidth}{0.3mm}}
 \vskip1ex}

\usepackage{color}
\definecolor{bluecolor}{rgb}{0,0.,1.}

\definecolor{redcolor}{rgb}{.7,0.,0.}

\usepackage{color}
\definecolor{bluecolor}{rgb}{0,0.,1.}

\definecolor{redcolor}{rgb}{.7,0.,0.}

\begin{document}
\title{Effect of noise in open chaotic billiards}
\author{Eduardo G. Altmann}
\email[Author to whom correspondence should be sent. E-mail address:
      ]{edugalt@pks.mpg.de}
\affiliation{Max Planck Institute for the Physics of Complex Systems, 01187 Dresden, Germany}

\author{Jorge C. Leit\~ao}
\affiliation{CFP and Departamento de F\'{i}sica, Faculdade de Ci\^{e}ncias Universidade do Porto, P-4169-007 Porto, Portugal}
\affiliation{Max Planck Institute for the Physics of Complex Systems, 01187 Dresden, Germany}

\author{Jo\~ao Viana Lopes}
\affiliation{CEsA - Centre for Wind Energy and Atmospheric Flows,\\
Faculdade de Engenharia da Universidade do Porto, 4200-465 Porto, Portugal }

\date{\today}

\begin{abstract}
We investigate the effect of white-noise perturbations on chaotic trajectories in open billiards. We focus on the temporal decay of the
survival probability for generic mixed-phase-space billiards. The survival probability has a total of five different decay regimes that
prevail 
for different intermediate times. 
We combine new calculations and recent results on noise perturbed Hamiltonian systems to characterize the origin of these regimes, and to compute how the parameters scale with noise intensity and billiard openness. Numerical simulations in the annular billiard support and
illustrate our results.
\end{abstract}

\maketitle

{\bfseries \noindent
In the last decade, experiments and applications strengthened the need of considering openness in otherwise closed billiards. 
The relevant dynamics in such systems is the {\it transient} emptying of the billiard, instead of time {\it asymptotic} properties of closed
systems. For instance, instead of the positive Lyapunov exponents of
typical trajectories, in open systems the signature of strong chaos is an exponential decay of the survival probability inside the
billiard. In this paper we 
perform a further step in this {\it asymptotic-to-transient} direction: we
put forward the idea that pre-asymptotic decay regimes of the survival probability are well defined and physically relevant.
We characterize the different regimes that dominate the intermediate time dynamics of noise-perturbed 
billiards, and we relate those regimes to the invariant structures of the closed phase space.
\spliter
}

\begin{figure*}[!ht]
\includegraphics[width=1\linewidth]{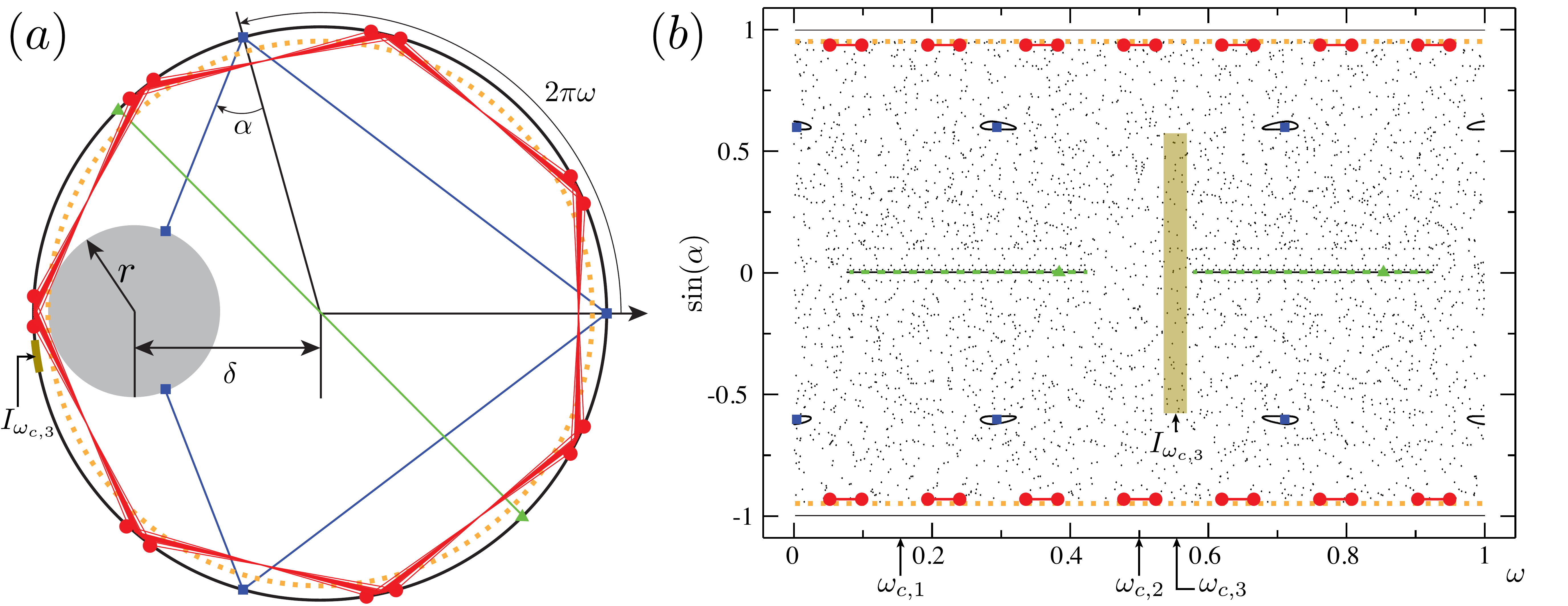}
\caption{(Color online) The annular billiard with~$r=0.3$ and $\delta=0.65$. (a) Configuration space showing selected periodic orbits. (b) Phase
  space obtained as a Poincar\'e surface of section at the outer circle. Shown MUPOs: 
$(2,1)$ (green triangles), $(7,1)$ (red circles) and KAM island around the period $6$ orbit (blue squares). Black dots in (b) correspond to a single chaotic trajectory.
Three positions of leaks are indicated: $\omega_{c,1} = 0.161$, $\omega_{c,2} = 0.5$, and $\omega_{c,3} = 0.55$ (the shaded region shows the
leak~$I_{\omega_{c,3}}$ used in Figs.~\ref{fig.2} and~\ref{fig.3}).} 
\label{fig.1}
\end{figure*}

\section{\label{sec.I} Introduction}

The idea of considering a hole or leak in an otherwise closed billiard is very appealing. This was the physical picture used by Pianigiani and
Yorke in their influential work on conditionally invariant measures~\cite{PY:1979}. Nowadays it is an essential element in the theoretical description
of experiments in atomic, acoustic, microwave, and optical cavities~\cite{Friedman,Kuhl2005,Harayama2011}. 

Another important ingredient considered in this paper is the effect of random independent and identically distributed perturbations (noise) on trajectories. There are many good reasons to consider it: test the  
structural stability of the results, facilitate calculations (e.g., random phase approximations), and mimic the effect of well-defined
physical processes that are too complicated or high-dimensional to be modeled in detail (e.g., molecular diffusion). 
Not surprisingly, the effect of noise is a classical problem in dynamical systems. In low-dimensional chaotic Hamiltonian systems
more than three decades of research considered the effect of noise on diffusion and anomalous transport~\cite{rechester,karney,floriani}, on the
trapping of trajectories~\cite{pogorelov,altmann.higherN,rodrigues,kruscha,kruscha2}, on scattering~\cite{seoane,altmann.noise}, etc. 

In this paper we consider the effect of white noise on the dynamics of open billiards. We are interested in the generic case of billiards
with mixed phase space in which regions of regular and chaotic motion coexist. We show how the survival probability of trajectories inside
the billiard is modified due to noise, and how the transition times and parameters of this modified survival probability scale with
noise intensity and leak size. The paper has a straightforward organization: we start with the closed billiard (Sec.~\ref{sec.II}),
which we leak (Sec.~\ref{sec.III}), and perturb by noise (Sec.~\ref{sec.IV}).

\section{\label{sec.II} Closed billiard}

\subsection{Definition of the dynamics}\label{ssec.IIdynamics}
The annular billiard is defined by two eccentric circles with ratio $0<r<1$ between the two radii and distance~$0 \le \delta < 1-r$ between
the two centers. Since its introduction by Sait\^o et al. three decades ago~\cite{saito}, the annular billiard has been used to investigate
different physical phenomena~\cite{bohigas,hentschel,egydio}. Here we consider the case~$r=0.3$ and $\delta=0.65$ shown in
Fig.~\ref{fig.1}(a). These parameters were chosen because numerical simulations shown in Fig.~\ref{fig.1}(b)
  strongly indicate a mixed phase space with the coexistence of regions of regular and chaotic motion. Below we discuss the main properties of this particular billiard 
but we emphasize the generality of our main results to the generic case of mixed-phase-space billiards.

The dynamics in the annular billiard can be easily constructed by noting that after colliding with the outer circle, trajectories can either collide directly with the outer circle or instead first collide once with the inner circle. Trajectories that collide with the  inner
  circle fulfill the following condition~\cite{saito}  
\begin{equation}\label{eq.collision}
|\sin\alpha + \delta\sin(\alpha-2\pi \omega)| \leq r,
\end{equation}
where $\alpha$ is the angle between the velocity and the normal vector at the collision point and $\omega$ is the normalized position of
collision at the outer circle (see Fig.~\ref{fig.1}a).
The significance of the hitting condition in Eq.~(\ref{eq.collision}) is that it can be used to obtain a mapping between two successive collisions at the outer
boundary~\cite{saito}.
We investigate this billiard map, which 
corresponds to a Poincar\'e surface of section at the outer boundary. For convenience, time is counted discretely between successive
collisions of the map. The collision time ($t_{col}$, assuming constant velocity~$v=1$) in the {\it chaotic component} of the annular billiard is bounded by
$\min\{1-r-\delta,2\sqrt{1-(r+\delta)^2}\}<t_{\text{coll}}< 2$, so that the scalings and the existence of regimes of decay remain unaffected
by this simplification\footnote{In ergodic closed systems the correspondence between discrete and continuous time is given by the mean
 collision time $\left\langle t_{\text{coll}}\right\rangle  = \frac{\pi A}{d v}$, where $A$ and $d$ are the area and perimeter of the billiard. In open systems this does not hold~\cite{mortessagne,altmann.rmp}. The specific
 quantitative results change differently in different parts of the survival probability, however, the scalings remain unchanged.}.

\subsection{Phase space components}\label{ssec.phasespace}

The phase space in Birkhoff coordinates $(\omega, \sin \alpha)$ of the closed annular billiard is shown in Fig.~\ref{fig.1}(b). 
It can be divided in four invariant components (i)-(iv), which are built by the trajectories that:

\begin{itemize}

\item[(i)] do not cross the inner circle of radius $r+\delta$ ($\sin \alpha \ge r+\delta$) and therefore never satisfy
  Eq.~(\ref{eq.collision}). Graphically they correspond to orbits close to the outer boundary of the circle and are called {\bf whispering 
gallery}. For the billiard used in this paper, the whispering gallery exists for $|\sin \alpha|<0.95$, beyond the dotted line in Fig.~\ref{fig.1}.

\item[(ii)] cross the circle of radius~$r+\delta$ ($\sin \alpha < r+\delta$) but never satisfy Eq.~(\ref{eq.collision}). These
  conditions are satisfied by periodic orbits that build one-parameter families of marginally unstable orbits (MUPOs)~\cite{Gaspard}. One trivial
  example is the diameter (period $p=2$ and winding number $q=1$ orbit) highlighted in Fig.~\ref{fig.1}. In opposite to the whispering
  gallery, these orbits are 
  usually embedded in a chaotic component
  (see below) and affect the dynamics of chaotic trajectories despite having zero measure, see Ref.~\cite{altmann.mupos} for a detailed investigation in the
  annular billiard and Refs.~\cite{Fendrik,Gaspard,Akaishi2009,OrestisMushroom,OrestisChaos} in other systems. For the
  billiard used in this paper the following MUPOs $(p,q)$ exist: $(2,1), (6, 1), (7, 1), (8, 1), (9, 1), (19, 2), (29, 3), (39, 4)$,
    $(49,5), (59, 6), (69, 7), (79, 8), (89, 9),\ldots$. 

\item[(iii)] cross the circle of radius~$r+\delta$ ($\sin \alpha < r+\delta$), satisfy Eq.~(\ref{eq.collision}), but remain
  close to stable a periodic orbit. These orbits can be periodic, quasi-periodic, or even chaotic (confined inside the last quasi-periodic
  circle) and build the so-called Kolmogorov-Arnold-Moser (KAM) islands.  For the billiard used in this paper the most prominent examples are the trajectories
  around the period $6$ orbit shown as $\blacksquare$ in Fig.~\ref{fig.1}.

\item[(iv)]  cross the circle of radius~$r+\delta$ ($\sin \alpha < r+\delta$), satisfy Eq.~(\ref{eq.collision}), are chaotic, and
  fill a large component of the phase space.
Despite the 
  mathematical difficulties to provide rigorous proofs (see Refs.~\cite{Chen,Foltin} for rigorous results in particular cases), it is largely believed that the annular billiard with $\delta\gg0$ contains a large
  chaotic component, in which a single trajectory visits a positive area of the phase space \cite{saito,bohigas,hentschel,egydio}. This large component is called {\bf chaotic
    sea}. In Fig.~\ref{fig.1}(b) this region corresponds to the large dotted component.
\end{itemize}

\begin{figure*}[!ht]
\includegraphics[width=\linewidth]{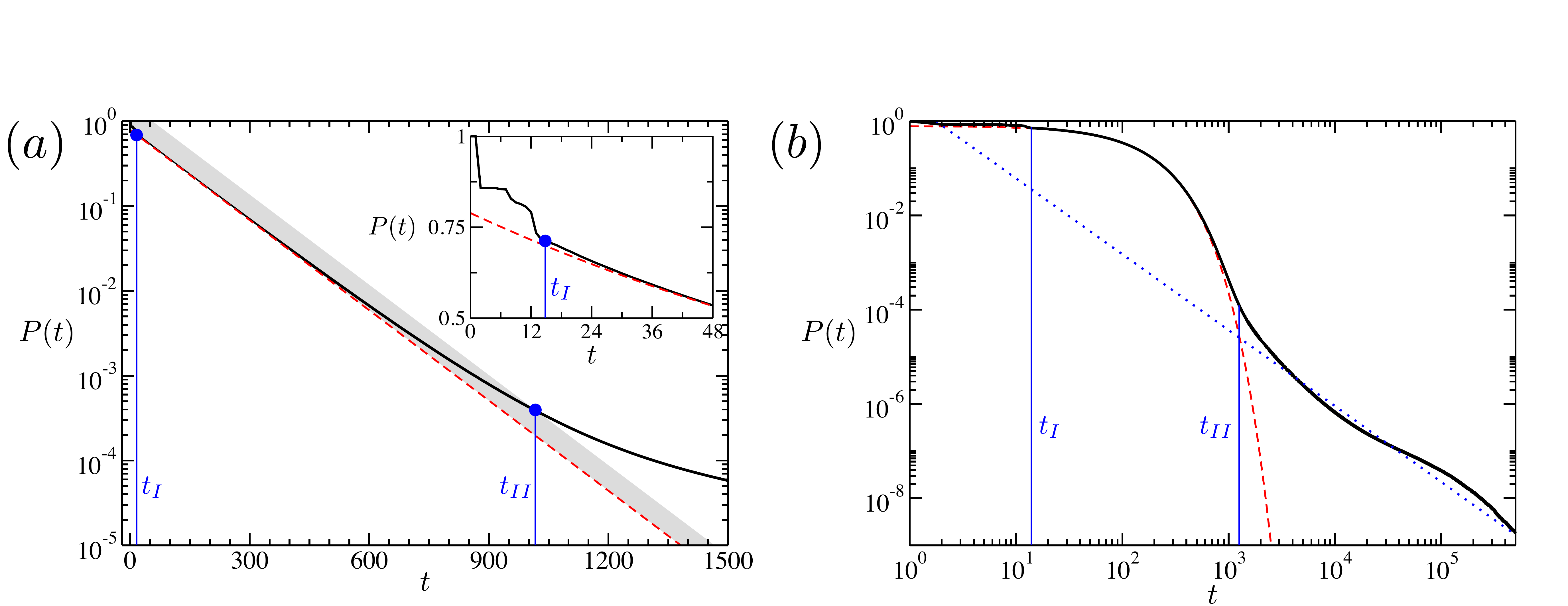}
\caption{(Color online) The survival probability~$P(t)$ for the open annular billiard. (a) Logarithmic scale
  in the $y$ axis; Inset: $x,y$ axes in linear scale, magnification for short times. (b)
  Logarithmic scale in both $x,y$ axes. All regimes and transition times in Eq.~(\ref{eq.deterministic}) are depicted: the dashed line
  corresponds to the fitting of an exponential with $a=0.72$ and $\gamma=0.016$; and the dotted line corresponds to the fitting of a
  power law with~$b=2.492$ and $\beta=1.65$. The transition times are $t_I=14$ (visual inspection) and $t_{II}=1,020$. The leak is introduced in the position $\omega_{c,3}$ with $\mu(I)=0.01$, see Fig.~\ref{fig.1}.}
\label{fig.2}
\end{figure*}

\section{\label{sec.III} Open billiard}

\subsection{Definition of the dynamics}

We open the annular billiard by considering a region~$I$ at the border of the billiard through which the trajectories escape~\cite{PY:1979,paar,Bunimovich,AltmannTel2,beims2010,OrestisStadium,OrestisChaos,altmann.rmp}. Formally, the dynamics of the system with leak~$\tilde{M}$ is defined based on the dynamics of the closed billiard~$M$ as:
\begin{equation}\label{eq.leak}
\tilde{M} = \left\{ \begin{array}{ll}
  \text{ escape }  & \text{ for } \vec{x} \in I, \\
  M & \text{ for } \vec{x} \notin I, \\
 \end{array} 
\right.
\end{equation}
where $\vec{x}=(\omega,\sin \alpha)$.
Notice that, by convention, escape occurs only one time step after trajectories hit the leak~$I$ so that~$\tilde{M}$ is defined in~$I$. 
 We are
interested in finite but small~$I$ such that a non-trivial dynamics still exists in the billiard.

Here we
consider leaks placed inside the large chaotic sea, item (iv) of Sec.~\ref{ssec.phasespace}, and we are interested on how the trajectories escape
from this ergodic component of the closed billiard's phase space.  The invariant set of the open system is the
so-called {\it chaotic saddle}~\cite{laitamas}, the set of points that never leave the system neither in forward nor in backward times.
In fully chaotic systems the chaotic saddle is a zero measure fractal set. In the case of mixed phase space system discussed here, the
chaotic saddle relevant to the escaping trajectories also contains a similar hyperbolic component~\cite{Jung,AltmannTel2} but additionally it includes a
non-hyperbolic component composed by the borders of the whispering gallery [region (i) of  Sec.~\ref{ssec.phasespace}], the MUPOs  [region (ii) of
  Sec.~\ref{ssec.phasespace}], and of the KAM islands [region (iii) of  Sec.~\ref{ssec.phasespace}]. 

In our simulations of the annular billiard we consider leaks~$I=[\omega_c-\Delta \omega, \omega_c+\Delta \omega] \times [-\Delta
  \sin\alpha, +\Delta \sin \alpha]$ centered at three different positions $\omega_c$ (see Fig.~\ref{fig.1}), varying $\Delta \omega$, and a fixed $\Delta
  \sin\alpha= 2/3$.
 Physically, this configuration corresponds to a dielectric billiard with refraction index $\eta=(\Delta \sin \alpha)^{-1}=1.5$ (glass) with a
 perfect mirror boundary everywhere except inside the leak, where trajectories escape for collisions below the critical angle $\alpha_c$
 with $\sin \alpha_c = 1/\eta$.

\subsection{Survival probability}\label{ssec.det.surv}

We compute the survival probability~$P(t)$ inside the billiard by starting an ensemble of trajectories distributed according to an
initial density~$\rho_0(\vec{x})$. In our simulations we consider $\rho_0(\vec{x})$ to be uniform inside the leak of the billiard and $0$
elsewhere (for the first iteration the closed billiard $M$ is used). 
Physically, these initial conditions correspond to throwing trajectories inside the billiard through the leak. Another
motivation for using this particular $\rho_0(\vec{x})$ comes from the fact that~$P(t)$ in this case corresponds exactly to the distribution of Poincar\'e recurrence times~\cite{AltmannTel2}. The
main decay regimes of $P(t)$ remain unaffected by this choice of $\rho_0(\vec{x})$, but values of the exponents and transition times may
change [see bullet items after Eq.~(\ref{eq.deterministic})].

Figure 2 shows the decay of the survival probability for the particular leak shown in Fig.~\ref{fig.1}. We can identify three different regimes of decay~\cite{Fendrik,AltmannTel2}
\begin{equation}\label{eq.deterministic}
P_{\text{deterministic}}(t) \approx \left\{ \begin{array}{ll}
  \text{ irregular } & \text{ for } t < t_I, \\
   a e^{-\gamma t} & \text{ for } t_I < t < t_{II}, \\
   b t^{-\beta} & \text{ for } t > t_{II}. \\
\end{array} 
\right.
\end{equation}
The dynamics of a typical trajectory escaping in each of the regimes in Eq.~(\ref{eq.deterministic}) can be associated to the phase space
structures as: 
\begin{itemize}

\item $t<t_I$ irregular: trajectories that collide only a few  times in the chaotic sea of the closed  billiard [region (iv) of Sec.~\ref{ssec.phasespace}], the spatial density of survival trajectories has
  not converged yet. The exact shape in this regime is extremely sensitive to~$\rho_0(\vec{x})$. 

\item $t_I<t<t_{II}$ exponential $e^{-\gamma t}$: trajectories explore the hyperbolic component of the chaotic saddle but escape before
  coming close to the border of the non-hyperbolic components [(i,ii,iii) of
    Sec.~\ref{ssec.phasespace}]. Analogous to the case of fully chaotic systems~\cite{laitamas}, the same exponent $\gamma$ is
  observed for different $\rho_0(\vec{x})$~\cite{AltmannTel2}.

\item $t>t_{II}$ power-law $t^{-\beta}$: trajectories get stuck close to the non-hyperbolic components of the saddle (classical references on such
  stickiness phenomena are Refs.~\cite{chirikov1,meiss.ott.physicaD,zaslavsky.pr}). The asymptotic exponent changes from $\beta$ to
    $\beta'=\beta-1$  for 
  $\rho_0(\vec{x})$ nonzero at the boundary of the non-hyperbolic components (e.g., if $\rho_0(\vec{x})$ is taken according to the
  Liouville measure restricted to the chaotic sea of the closed billiard)~\cite{meiss,laitamas}.
\end{itemize}
The description above applies to typical trajectories escaping in the corresponding regimes. It is instructive to think that for
intermediate times, $t_\beta<t< t_{II}$, both exponential and power-law regimes
coexist~\cite{AltmannTel2,Akaishi2009,OrestisStadium}  
\begin{equation}\label{eq.coexist}
P(t) =   a e^{-\gamma t} + b t^{-\beta},
\end{equation}
where $t_{\beta}$ is the time needed to approach  the non-hyperbolic component of the saddle~\cite{AltmannTel2}.

\begin{figure*}[!ht]
\includegraphics[width=\linewidth]{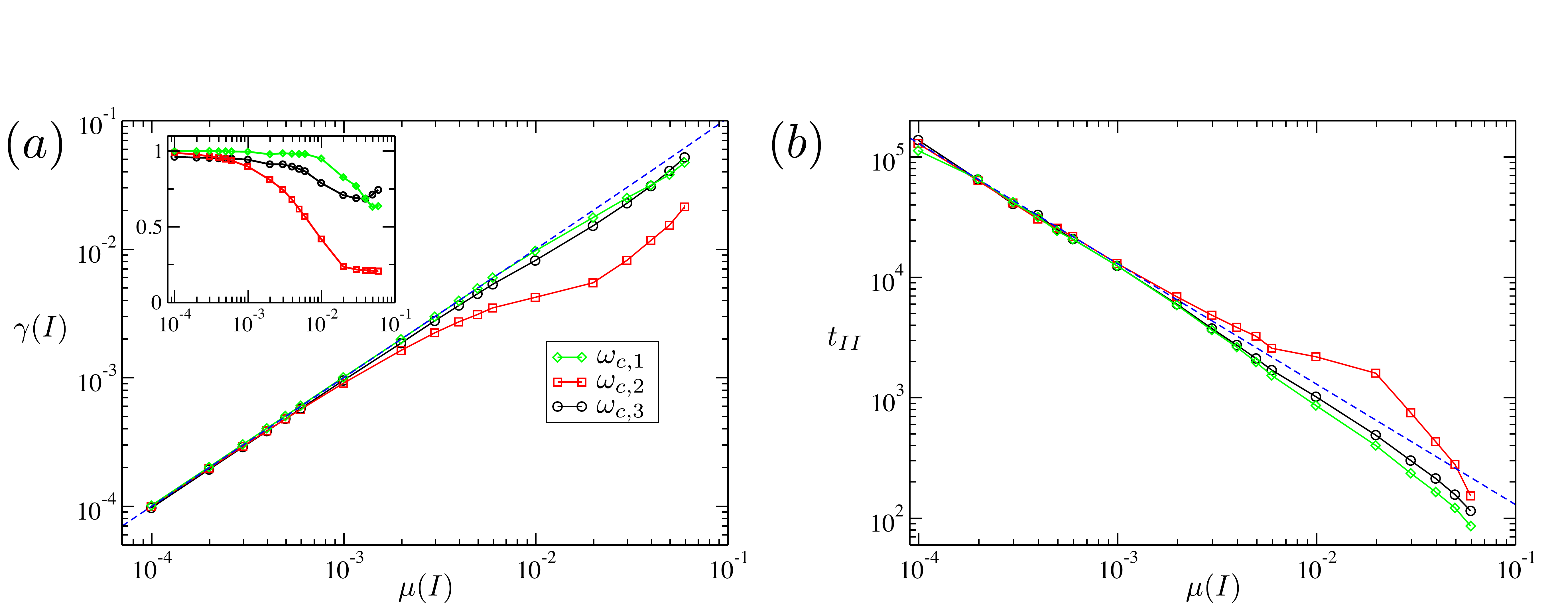}
\caption{(Color online) Scalings of the parameters of $P(t)$ in Eq.~(\ref{eq.deterministic}) with leak
  size~$\mu(I)$ for three different leak locations.
 (a) Intermediate times escape rate~$\gamma$; The dashed line corresponds to $\gamma^*$
  in Eq.~(\ref{eq.mustar}). Inset: dependence of the coefficient $a$ with $\mu(I)$. (b) Transition time $t_{II}$, obtained
  using Eq.~(\ref{eq.defnc}); the dashed line corresponds to a scaling $1/\mu(I)$.
The leaks have a fixed height~$\Delta \sin \alpha = 2/3$ and varying width $\Delta \omega \rightarrow 0$, which leads to $\mu(I)\rightarrow0$.
The centers of the leaks are: $\omega_{c,1} = 0.161$ (in a MUPO, black circles),
$\omega_{c,2}=0.5$ (in an unstable periodic orbit, red squares), and 
$\omega_{c,3}=0.55$ (in the chaotic sea, green diamonds). 
}
\label{fig.3}
\end{figure*}

\subsection{Dependence of the parameters of $P(t)$ on the leak}

The {\bf exponential decay~$\gamma$} in Eq.~(\ref{eq.deterministic}) can be considered a signature of the chaoticity of the map. Following
this reasoning, for small leaks  we can approximate the escape at each time step by the area of the leak relative to the area of the
chaotic sea $\mu(I):=\text{Area}(I)/\text{Area(chaotic sea)}$. For the billiard considered here we found numerically that $\text{Area(chaotic sea)} \approx
0.993\times(r+\delta) =  0.943$. Using phase-space areas correspond to using the Liouville measure $d\mu=d\omega d \sin \alpha$ of the closed  
system to approximate properties of the open system, and can be shown to be valid for almost all leak positions in
strongly chaotic systems~\cite{altmann.rmp}. This leads to an estimation of the exponential decay as
\begin{equation}\label{eq.mustar}
\gamma^*=-\ln(1-\mu(I)) \approx \mu(I) \text{ for } \mu(I)\rightarrow0.
\end{equation}
Violations of this approximation in the fully chaotic case have been extensively discussed in the recent years and are particularly large
for leaks containing low-period periodic orbits of the closed system~\cite{paar,AltmannTel2,Bunimovich,Demers}. Here we extend these previous
results and verify the
effectiveness of the approximation in Eq.~(\ref{eq.mustar}) for the intermediate-time exponential decay in Eq.~(\ref{eq.deterministic}). In Fig.~\ref{fig.3}(a) we
compare the numerically obtained values to the prediction for different leak sizes centered at three different positions: in the chaotic region, around
an unstable periodic orbit, and around a family of MUPOs. In all cases $\gamma \rightarrow \gamma^*$ is observed in the limit of small
leaks $\mu(I)\rightarrow0$, in agreement with relation~(\ref{eq.mustar}). For large leak sizes Fig.~{fig.3}(a) shows different deviations of this relation, in agreement with the results observed for hyperbolic systems~\cite{paar,AltmannTel2,Bunimovich,Demers}.

The importance of the value of~$\gamma$ is that the same value is obtained for a broad class of smooth initial densities $\rho_0$. In fully chaotic systems the requirement is
that $\rho_0$ intersects the stable manifold of the chaotic saddle. Analogously, the requirement here is that it intersects the stable manifold of the hyperbolic component of the chaotic saddles.

The {\bf power-law exponent~$\beta$} depends on the properties (of the boundary) of the non-hyperbolic sets embedded in the chaotic
component of the phase space (components (i), (ii) and (iii) in the list of Sec.~\ref{sec.II}). If no KAM islands are present, an exponent $\beta=2$ can be obtained for MUPOs~\cite{Fendrik,OrestisStadium,OrestisChaos}. The question of whether the asymptotic regime has a
well-defined and universal power law in the generic KAM case is still under investigation for the case of area-preserving maps (see Refs.~\cite{cristadoro,venegeroles} for the latest results
that indicate $\beta \approx 1.57$). For simplicity we write the asymptotic decay as $t^{-\beta}$, but it is meant to describe the
power-law like behavior usually observed in mixed-phase-space systems~\cite{chirikov1,meiss.ott.physicaD,zaslavsky.pr}.

\subsection{Dependence of the transition times of $P(t)$ on the leak}

Transition time $t_I$ indicates the starting of the exponential decay. It can be interpreted as a convergence time which is proportional
to~$1/|\lambda'|$,   where~$\lambda'$ is the  negative Lyapunov exponent of the saddle (the time to relax
  to the hyperbolic component of the saddle along its stable manifold). Numerical observations usually show an abrupt approach, i.e., the
  exponential decay provides a good description of $P(t)$ after a finite (short) time~\cite{AltmannTel2}.

The transition time $t_{II}$ is defined from Eq.~(\ref{eq.coexist}) as the time for which the exponential and power-law contributions are
equal~\cite{AltmannTel2} 
\begin{equation}\label{eq.defnc}
 a e^{-\gamma t_{II}}= b t_{II}^{-\beta} \Rightarrow p(t_{II})= 2 \gamma a e^{-\gamma  t_{II}}.
\end{equation}
The ratio $a/b$ can be interpreted as the proportion between the number of trajectories escaping exponentially to the number of trajectories
escaping algebraically. It depends mainly on the measures of the chaotic and regular components of the phase space and therefore it should not
depend strongly on the measure of the leak $\mu(I)$. Under this assumption we can estimate the scaling of $t_{II}$ on the leak size $\mu(I)$ as~\cite{AltmannTel2}
\begin{equation}\label{eq.tii}
t_{II} \sim \frac{1}{\gamma} \sim \frac{1}{\mu(I)},
\end{equation}
for which additional logarithmic corrections apply~\cite{Akaishi2009}. The scaling in Eq.~(\ref{eq.tii}) has been confirmed in our numerical
simulations for the three different leak positions, see Fig.~\ref{fig.3}(b).

\section{\label{sec.IV} Open noisy billiard}

\subsection{Definition of the dynamics}

Here we consider additive noise perturbations to the trajectories. In the simulations of the annular billiard we have implemented at each
collision a perturbation to the angle~$\alpha$ as
$$ \alpha'=\alpha+\delta,$$
where $\delta$ is an independent normal distributed random variable with zero mean, $\langle \delta \rangle=0$, and standard deviation
$\sigma=\pi \xi$ (noise strength). 
In order to prevent the particle from leaving the billiard through the border (non-physical situation), the noise distribution was truncated at $\alpha=\pm \pi/2$. Notice that perturbations in the $\alpha$ direction are perpendicular to the border of the whispering gallery component and to the parameterization of the billiard boundary, having therefore a strong
  impact on sliding orbits  ($\alpha=\pm \pi/2$)\footnote{In fact, trajectories have a tendency of being repelled
    from sliding orbits because the noise perturbation in $\alpha$ is nonzero, truncated in $\alpha = \pm \pi/2$, and added on
    each collision. Therefore, we do not expect these orbits to dramatically affect $P(t)$ or the relation between billiard maps and flows in our case.}.
%
Based on previous observations with different setups~\cite{altmann.higherN,altmann.noise,kruscha,kruscha2,rodrigues}, and in the generality of the
arguments below, we believe our results are valid for additive white noise in general (provided the perturbation in $\alpha$ is
  nonzero).  It is an interesting open question whether (and which) modifications are needed for multiplicative and colored noise (see Ref.~\cite{pogorelov}).

\begin{figure*}[!ht]
\includegraphics[width=\linewidth]{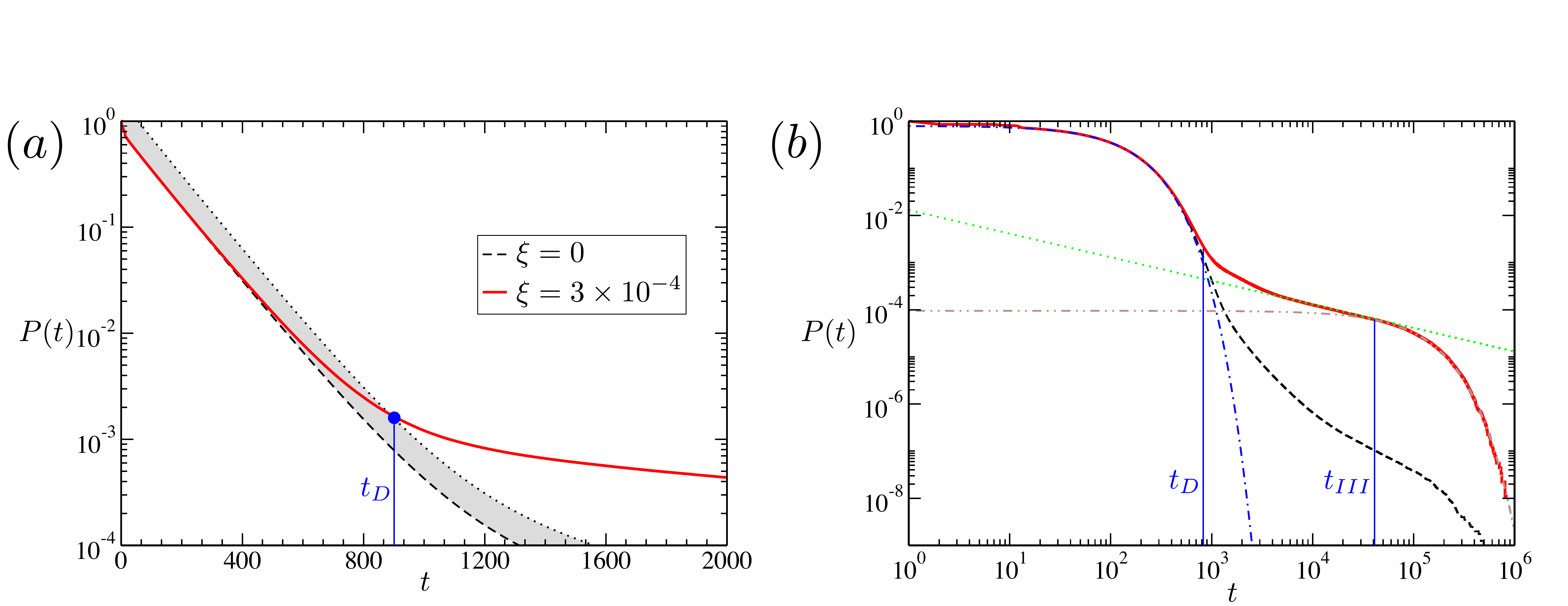}
\caption{(Color online) Survival probability~$P(t)$ for the open annular billiard perturbed by noise with intensity~$\xi=3\times 10^{-4}$. (a) Logarithmic scale in the y axis; (b)
  Logarithmic scale in both x and y axis. All regimes and transition times in Eqs.~(\ref{eq.both})-(\ref{eq.noise}) are depicted in (b):
  fitting of the asymptotic exponential with $d=9.34\times 10^{-5}$ and $\gamma_\xi=1.06\times 10^{-5}$ (dot-dashed line in brown);
  power-law decay with~$c=0.014$ and $\beta_{RW}=0.5$ (dotted line in green). The dashed black line corresponds to the $\xi=0$ case (see
  Fig.~\ref{fig.2}). The transition times were estimated as $t_D=887$ and $t_{III}=15,158$. The leak is as in Fig. \ref{fig.2}
  ($\omega_{c,3}$ with $\mu(I) = 0.01$).}  
\label{fig.4}
\end{figure*}

\subsection{Survival probability}

In Sec.~\ref{ssec.det.surv} the trapping of trajectories inside the billiard was connected to invariant structures of the
deterministic phase space. The longer the escape time of trajectories, the closer they approach these invariant structures. This leads to a
connection between temporal scales of the survival probability and spatial scales in the phase space. Noise perturbations affect phase-space scales comparable to~$\xi$. Based on these arguments we expect that for small~$\xi$ the survival probability~$P(t)$ is
modified for long times only:

\begin{equation}\label{eq.both}
P(t) \approx \left\{ \begin{array}{ll}
  P_{\text{deterministic}} & \text{ for } t < t_D, \\
  P_{\text{noise}} & \text{ for }  t > t_D , \\
 \end{array} 
\right.
\end{equation}
where $P_{\text{deterministic}}$ is given by Eq.~(\ref{eq.deterministic}) and $t_D$ is the transition time. Following Refs.~\cite{altmann.higherN,altmann.noise}, the
noise perturbed survival probability is given by

\begin{equation}\label{eq.noise}
P_{\text{noise}}(t) \approx \left\{ \begin{array}{ll}
  c t^{-\beta_{RW}} & \text{ for } t_{D} < t < t_{III}, \\
  d e^{-\gamma_\xi t} & \text{ for } t > t_{III}. \\
 \end{array} 
\right.
\end{equation}
Figure~\ref{fig.4} shows, for $\xi=3\times10^{-4}$, the different decay regimes and transition times of the survival probability given by Eqs.~(\ref{eq.both})-(\ref{eq.noise}). 
The dynamics of a typical trajectory escaping in each of these regimes
can be associated to the phase space structures of the deterministic dynamics as: 
\begin{itemize}

\item $t<t_D$ deterministic: trajectories escape before the noise perturbation is noticed, $P(t)$ coincides with the $\xi=0$ case.

\item $t_{D} < t < t_{III}$ enhanced trapping: trajectories enter the region corresponding to regular motion of the deterministic dynamics
  [components (i) and (iii) of Sec.~\ref{ssec.phasespace}] and perform a one-dimensional random walk inside it~\cite{altmann.higherN,rodrigues,altmann.noise,kruscha,kruscha2}. 

\item $t>t_{III}$ asymptotic exponential: trajectories explored all available phase-space.

\end{itemize}
In this description we neglect the effect of noise on~$\gamma$, which has been investigated for a fully chaotic system in
Ref.~\cite{altmann.noise}. The sticky region around MUPOs [region (ii) of Sec.~\ref{ssec.phasespace}] does not contribute to the enhanced
trapping regime ($t_{D} < t < t_{III}$) because MUPOs build a zero measure set.  This means that if noise is added to a system in which
MUPOs are the only source of stickiness (e.g., the Stadium~\cite{OrestisStadium} or Drive-belt billiards~\cite{OrestisChaos}), we predict
that the exponential decay will start immediately after $t_{D}$ (i.e., $t_{III}=t_{D}$).

\begin{figure*}[!ht]
\includegraphics[width=\linewidth]{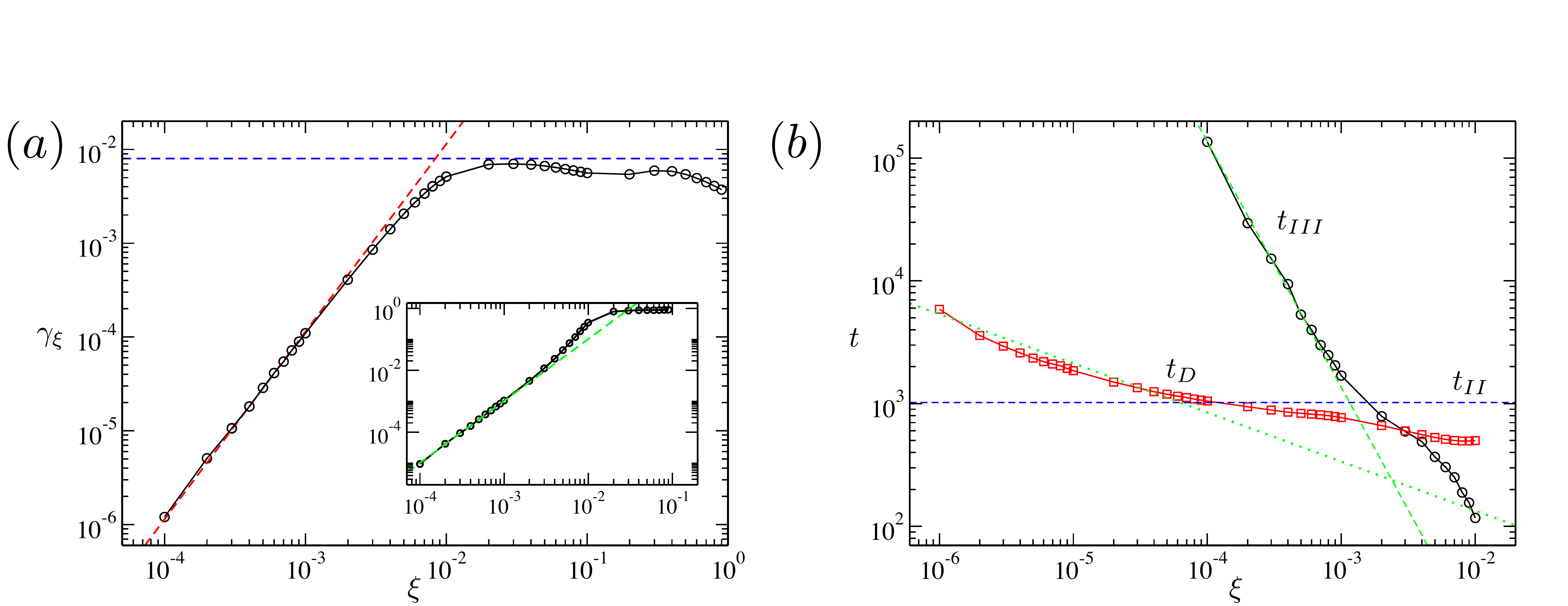}
\caption{(Color online) Dependence of the parameters of $P(t)$ in Eq.~(\ref{eq.noise}) with noise intensity~$\xi$.  (a)
  Fitted $\gamma_\xi$ with scaling $\xi^2$ (red dashed line) and value $\gamma$ (blue dots).  Inset: coefficient $d$ vs. $\xi$ with
  green dashed line representing $\xi^{-2}$; (b) $t_{III}$ 
  obtained as the time for which $P(t)$ intersects the fitted curve $1.1 d\exp(-\gamma_\xi t_{III})$ (black circles); $t_D$ obtained as the time
  for which $P(t)$ intersects $2 P_{\text{deterministic}}(t)$ (red squares); and  green dotted lines indicating scalings $\xi^{-0.4}$
  (bottom) and  $\xi^{-2}$ (top). The leak is the same as in Figs.~\ref{fig.2} and \ref{fig.4}. }
\label{fig.5}
\end{figure*}

\subsection{Dependence of the parameters of $P(t)$ on the leak}

For small noise perturbations the parameter $\beta_{RW}$ in Eq.~(\ref{eq.noise}) can be related to the scaling of the recurrence time
distribution of a one-dimensional random walk (with step size $\sim \xi$) and is therefore given by~\cite{Feller,altmann.higherN}
\begin{equation}\label{eq.betarw}
\beta_{RW}=\frac{1}{2} \text{ for } \xi\rightarrow 0. 
\end{equation}
In the derivation of this result in Ref.~\cite{altmann.higherN} (see also~\cite{altmann.noise}) the initial conditions~$\rho_0(\vec{x})$ were
  chosen {\it outside} KAM islands of the deterministic closed billiard, in agreement with the case treated here. See
  Refs.~\cite{rodrigues,kruscha,kruscha2} for the case of $\rho_0(\vec{x})$ {\it inside} KAM islands. References~\cite{kruscha,kruscha2}
also showed that if $\xi^2$ terms are included, the random walk is biased.

The scaling in Eq.~(\ref{eq.betarw}) is interrupted for long times because of the limited region available for the random walk in the KAM islands and whispering gallery. In the random-walk model this corresponds to adding a reflecting boundary condition~\cite{altmann.higherN,altmann.noise}. The exponent~$\gamma_{\xi}$ of the asymptotic decay can be obtained
considering $P(t)$ to be a continuous and smooth function around $t=t_{III}$. Evaluating
\begin{equation}\label{eq.gammaxi1}
\frac{\partial \log(P(t))}{\partial t} \text{ at } t=t_{III}
\end{equation}
for both terms in Eq.~(\ref{eq.noise}) and equating then leads to $t_{III}=0.5/\gamma_\xi$. Below we show that
$t_{III}\sim1/\xi^2$ [see Eq.~(\ref{eq.tiii}) and Refs.~\cite{altmann.higherN,altmann.noise}], and therefore we obtain
\begin{equation}\label{eq.gamma_xi}
\gamma_\xi \sim \frac{1}{t_{III}} \sim \xi^{2} \text{ for } \xi\rightarrow 0.
\end{equation}
This scaling is confirmed in Fig.~\ref{fig.5}(a). Interestingly, we observe that for larger noise intensities $\gamma_\xi$ experiences a
crossover to $\gamma_\xi\approx \gamma$, i.e., $\gamma_\xi$ is bounded by the escape rate observed for short times which was related to the hyperbolic
component of the chaotic saddle in Sec.~\ref{sec.II}. This observation indicates that once the
trajectories leave the region corresponding to regular motion of the deterministic dynamics, the hyperbolic component of the
saddle controls their escape. For small noise the process of leaving the regular components is slower and therefore the
scaling~(\ref{eq.gamma_xi}) dominates~$\gamma_\xi$. For larger noise, the deterministic
exponential escape is slower than the escape from the islands and therefore $\gamma_\xi\approx\gamma$ is observed. In Ref.~\cite{rodrigues} a
similar but different scaling of $\gamma_\xi$ on $\xi$ was numerically obtained for the case of random maps and for initial conditions
taken inside the KAM islands.

\subsection{Dependence of the transition times of $P(t)$ on the leak}

A theory for $t_D$ can be found in Ref.~\cite{floriani} and predicts that
\begin{equation}\label{eq.tD}
t_D \sim 1/t^\Lambda \text{ for } \xi\rightarrow0,
\end{equation}
with $\Lambda\lessapprox 1$ related to the scaling of {\it Cantori} close to the KAM islands. This scaling is valid for $t_D \gg t_{II}$ because
for $t_D \simeq t_{II}$ the deterministic trapping around the KAM islands has a limited contribution. 

The final cut-off time~$t_{III}$ can be estimated from basic properties of diffusion motion. The expected distance~$L$
traveled by a random walker with step-size $\sim \xi$ grows as $L \sim \xi \sqrt{t}$. The time $t_{III}$ corresponds to the 
expected time for the $\xi$-perturbed trajectories to travel the (fixed) distance corresponding to the diameter of the (largest) KAM
island. Therefore, we estimate
\begin{equation}\label{eq.tiii}
t_{III} \sim \xi^2, \text{ for } \xi\rightarrow 0.
\end{equation}

Our numerical simulations for the scaling of the transition times are shown in Fig.~\ref{fig.5}(b). We see that for times comparable
to~$t_{II}$ only a weak dependence of $t_D$ on $\xi$ is observed. For times larger than $t_{II}$ our results indicate an increase of this
dependency, consistent with relation~(\ref{eq.tD}). The scaling of $t_{III}$ in Eq.~(\ref{eq.tiii}) is confirmed over a larger interval of $\xi$.

\section{\label{sec.V} Summary of Conclusions}

White noise perturbations have surprising effects on the chaotic dynamics of mixed-phase-space Hamiltonian
systems~\cite{rechester,karney,floriani,pogorelov,altmann.higherN,rodrigues,kruscha,kruscha2,seoane,altmann.noise}. Here we have shown how
these results impact the dynamics of open billiards, in which case the openness can be controlled systematically and the phase space
  usually contains MUPOs~\cite{Gaspard,altmann.mupos}. We have combined these results into a complete survival probability which is obtained introducing Eqs.~(\ref{eq.deterministic})
and~(\ref{eq.noise})  
into Eq.~(\ref{eq.both}). For small noise intensities, it contains five different regimes with four transition times $(t_I,t_{II}, t_{III},t_D)$
and four physically relevant parameters $(\gamma,\beta,\beta_{RW},\gamma_\xi)$. We have discussed how these quantities depend on the
intensity of the perturbation~$\xi$ and on the position and size~$\mu(I)$ of the leak.

Apart from extending previous results to billiard systems, our paper contains several new findings. First, we have shown that the (intermediate-times) escape rate scales linearly with
leak size and is extremely sensitive to the location of the leak. These results were previously known for fully chaotic systems~\cite{paar,AltmannTel2,Bunimovich,Demers}. Second, we have
shown that the asymptotic exponential decay~$\gamma_\xi$ depends on the noise intensity $\xi$ as $\gamma_\xi \sim \xi^2$, with a transition
towards $\gamma_\xi = \gamma$ for large $\xi$ (Fig.~\ref{fig.5}a). Altogether, our results further emphasize the significance of the intermediate time decay
regimes of the survival probability in weakly chaotic noise-perturbed billiards.

\begin{acknowledgments}
 We thank T. T\'el for insightful discussions that led to Fig.~\ref{fig.5}(a), O. Georgiou and M. Matos for the careful reading of the
 manuscript. J. C. Leit\~ao acknowledges funding from Erasmus N 29233-IC-1-2007-1-PT-ERASMUS-EUCX-1.

\end{acknowledgments}
%

\end{document}